# Supercurrent interference and its transfer in a kagome superconductor

Heng Wu[1,2,3,*], Houssam el Mrabet Haje[1,2], Michiel Dubbelman[1,2], Brenden R. Ortiz[4], Stephen D. Wilson[5], Mazhar N. Ali[1,2], Yaojia Wang[1,2,3,*]

[1]Department of Quantum Nanoscience, Faculty of Applied Sciences, Delft University of Technology, Lorentzweg 1, 2628 CJ, Delft, The Netherlands

[2]Kavli Institute of Nanoscience, Delft University of Technology, Lorentzweg 1, 2628 CJ, Delft, The Netherlands

[3]Quantum Solid State Physics, KU Leuven, Leuven 3001, Belgium

[5]Materials Science and Technology Division, Oak Ridge National Laboratory, Oak Ridge, Tennessee 37831, United States

[4]Materials Department, University of California Santa Barbara, Santa Barbara, California 93106, U.S.A.

*e-mail: wuhenggcc@gmail.com, yaojia.wang@kuleuven.be

## Abstract

Superconductivity represents a macroscopic quantum state notable for its rich manifestations of electronic coherence and collective behavior. Kagome materials $AV_3Sb_5$ (A= K, Cs, Rb) possess cascade intertwined quantum phases including superconductivity, symmetry-breaking charge orders, nematic orders and topological states, making them attractive materials for exploring exotic superconducting states. However, the superconducting properties and the Cooper pairing behaviors have not been fully explored and understood. In this work, by studying both the magnetoresistance and critical current behaviors in $KV_3Sb_5$ ring and pristine flakes, we reveal the charge 2$e$ paring in $KV_3Sb_5$ although anomalous oscillations with smaller periodicity were observed, and report the intrinsic superconducting phase coherence in $KV_3Sb_5$ flakes. The former is demonstrated by the careful verification of the Little-Parks oscillations in differential resistance colormaps, and the latter indicates the existence of superconducting domains in $KV_3Sb_5$. Moreover, we observed a special phenomenon: the transfer of supercurrent interference patterns between the superconducting ring and the superconducting flake, which demonstrates the global critical current effect of the superconducting phase coherence. These findings provide new insights into the Cooper pairing behaviors in $KV_3Sb_5$ and highlight the importance of global effect of superconducting phase coherence in the understanding of the superconducting behaviors.

## Introduction

Superconductivity emerges when electrons form Cooper pairs, enabling dissipationless current flow and coherent superconducting phenomena such as flux quantization and the Josephson effect. The Josephson effect occurs in weak links (for example insulators, normal metals or geometrical constrictions) between superconductors, where a finite supercurrent can flow across the barrier [1]. Flux quantization is a hallmark of superconducting loops, where the phase winding of the order parameter around the loop is quantized, giving rise to phenomena such as the Little-Parks effect, showing periodic oscillations of the superconducting transition temperature $T_c$ with flux [2]. The behavior of phase coherent phenomena is closely related to the pairing mechanism in the devices. In conventional spin-singlet superconductors, where Cooper pairs consist of two electrons with opposite spins, the fundamental flux quantum and Josephson phase periodicity are set by the charge 2$e$, yielding a flux periodicity $\Phi_0=h/2e$ with h being the Planck constant and a 2π phase winding. These characters can differ in unconventional systems. For instance, Josephson junctions or rings with spin-triplet superconducting pairing (electrons with equal spins pairing together) may induce phase shifts [1,3,4], and an emergent concept of superconductivity with multicharge pairing (e.g., 4$e$, 6$e$) was proposed [5–9], which may lead to flux periodicities $\Phi_0=h/ne$ with n being the number of paired electrons. While Bardeen-Cooper-Schrieffer (BCS) theory successfully accounts for conventional spin-singlet pairing, the nature of unconventional superconducting phenomena remains an active topic of investigation.

Kagome material $AV_3Sb_5$ (A = K, Cs or Rb) is a family possessing multiple intertwining quantum states including superconductivity, topology, and time-reversal and rotational symmetry breaking charge orders (e.g. chiral charge order, nematic orders) [10–18]. These intertwined phases raise the possibility of exotic superconducting states, such as unconventional superconducting states [19,20], pair density waves with spatially modulated superconducting order parameters [21–24] and chiral superconductivity with time-reversal symmetry breaking [23,25,26]. Notably, Fraunhofer patterns and Little-Parks-like oscillations modulated by thermal cycling have been observed in intrinsic $CsV_3Sb_5$ flakes, which are believed to be related to the dynamic superconducting domains and potential chiral superconductivity [27,28]. More strikingly, based on the measurement of magnetoresistance oscillations, the exotic 4$e$ and 6$e$ pairing were reported in superconducting ring of $CsV_3Sb_5$ [29]. However, the nature of Cooper pair paring symmetry in this kagome family remains unsettled and under debate [30–33]. Given by the intertwined quantum phases and fertile superconducting states, it is of critical importance to study the superconducting phase coherence in detail for deepening the understanding of Cooper pair and exploring intriguing superconducting phenomena.

In this work, we report intrinsic superconducting interference phenomena in pristine $KV_3Sb_5$ flakes and demonstrate an unexpected transfer of supercurrent interference from a superconducting ring to the adjacent pristine flake. Magnetoresistance on a $KV_3Sb_5$ ring shows oscillations with the conventional period corresponding to $\Phi_0=h/2e$ at low temperature, but we also observe anomalous oscillations with smaller period at higher temperature. Detailed differential resistance measurement reveals that these anomalous oscillations are not consistent with the conventional Little-Parks oscillations. Detailed comparison between ring and superconducting flake reveals that they arise from intrinsic supercurrent interference of $KV_3Sb_5$ flake, which should be associated with superconducting domains. Furthermore, by changing the measurement configuration, we show that Little-Parks oscillations of the superconducting ring can be transferred to the adjacent flake within the current source path, and the critical current of the flake shows a current-polarity dependent behavior. These results show the unprecedented transfer of supercurrent interference between different superconducting regions, highlighting the global critical current effect of superconducting phase coherence in $KV_3Sb_5$ devices and underscore the importance of complementary measurements (including both magnetoresistance and critical current, as well as different measurement configurations) when interpreting fractional or anomalous flux periodicities in complex quantum materials.

## Results

$KV_3Sb_5$ crystalizes in the hexagonal structure with space group of P6/mmm, where V atoms form the kagome lattice, as shown in figure 1a. The single crystal can be exfoliated down to thin films due to weak bonds between the alkali layer and the Sb layer. We first fabricated the superconducting ring structure to investigate the behavior of Little-Parks effect. The inset of figure 1b shows the superconducting device (device #1) of a square ring, for which the side lengths of the inner square and the outer square of the ring are 0.63±0.02 µm and 1.25±0.01 µm, respectively. The detailed method for the device fabrication can be found in the Supplementary Materials. A standard 4-probe method is used to measure the resistance ($R$) of the ring, and the $R$-$T$ curve is shown in Fig. 1b. In most studies on the superconducting ring, oscillations on magnetoresistance ($R$ vs magnetic field ($B$)) are widely used to characterize the Little-Parks effect, as it is much easier to be measured than the oscillations of $T_c$. Figure. 1c shows the $R$-$B$ curves at different temperatures, where oscillations of the magnetoresistances can be clearly observed in all curves. Notably, the oscillations show different periodicity when increasing temperatures. We used the Fast Fourier Transition (FFT) method to analyze the oscillations of magnetoresistance. The detailed process of the FFT analysis can be found in the Supplementary Material. At low temperature, such as 0.24 K, the *R-B* curve shows periodic oscillations with the frequency of 439 (1/T) (Fig. 1e), corresponding to periodicity of $\Delta B$ = 2.3 mT. Based on the flux quantization with $\Phi_0 = h/2e = \Delta B \cdot S_{eff}$, the effective ring area $S_{eff}$ is 0.9 µm$^2$, which is in between the inner area (~0.40 µm$^2$) and the outer area (~1.56 µm$^2$) of the superconducting ring, and corresponds to a square whose side (length of ~0.95 µm) lies midway between the inner and outer edges. Therefore, we confirm that this set of oscillations in the *R-B* curve originates from the Little-Parks effect with conventional charge 2*e* pairing. Additionally, the oscillation is weak at 0.2 K and becomes stronger when the temperature is near $T_c$. This also consistent with the behavior of Little-Parks effect, as it is more obvious when the temperature is near $T_c$.

When further increasing the temperature, the Little-Parks oscillations become weaker and new oscillations with smaller periods emerge and dominate the oscillations on *R-B* curves at high temperature. As an example, the FFT result of the magnetoresistance at 0.36 K is shown in Fig. 1f and 1g (detailed FFT analysis in Supplementary Materials). However, there are lots of peaks that cannot give a clear periodicity since the FFT results show comparable amplitudes at various frequency peaks. The question is whether these anomalous oscillations are from Little-Parks effect, which directly influences the understanding of Cooper pairing.

In order to rigorously investigate the origin of these anomalous oscillations in the magnetoresistances, we measured the differential resistance as a function of bias current and magnetic field at different temperatures, also called supercurrent interference pattern. The results are plotted as the colormaps shown in Fig. 2. Surprisingly, in the colormap at low temperature (0.2 K) shown in Fig. 2a, the supercurrent interference patterns exhibit several sets of oscillations. The set of oscillations that sits lower than 0.3 µA exhibit a slow height decay and very clear periodic behavior with a period of 2.27 mT, same period as that obtained from the *R-B* curve, which originates from the Little Parks effect with charge 2*e*. However, the field dependent critical current of other patterns shows either diamond shapes that is a typical behavior of a superconductor, or various oscillations on top of them. These anomalous oscillations are too complicate to be extracted and analyzed, just like the oscillations in the *R-B* curve at high temperatures. But the fact we could conclude is that they are irrelevant to the Little-Parks effect of the superconducting ring, since the periodic Little-Parks oscillation has already disappeared at 3.2 K, whereas the anomalous oscillations on top of the diamond shaped background remain (Fig. 2b). While it is reported that there exist 4*e* and 6*e* paring in CsV$_3$Sb$_5$ ring [29] which is reflected by the oscillations in the magnetoresistance at temperatures near and above $T_c$, it is not the case in our KV$_3$Sb$_5$ ring.

These results indicate that the anomalous oscillations may originate from the intrinsic supercurrent interference of the KV$_3$Sb$_5$ sample. To examine this, we fabricated another KV$_3$Sb$_5$ ring device (device #2) which allows the comparison of KV$_3$Sb$_5$ ring and strip segments. In this device (image shown in the inset of Fig.3a and 3b), the ring is a circle with an inner diameter of 0.84±0.02 µm and an outer diameter of 1.32±0.02 µm, which is connected to a strip segment. We fabricated multiple pairs of electrodes on the device, enabling the four-probe measurement on both ring and strip segments of the device. The $T_c$ of the ring segment can be extracted from the *R-T* curve shown in Supplementary Material Fig. S1, which is 0.58 K. Figure 3a shows the *R-B* curves of the ring segment at different temperatures (the measurement configuration is shown in the inset). All the curves exhibit clear oscillations with similar periodicity. The FFT analysis for the oscillations at *T* = 0.52 K shows one prominent peak located at 501 (1/T) corresponding to an oscillation periodicity of $\Delta B$ = 2.00 mT. The calculated effective ring area $S_{eff}$ based on $\Phi_0 = h/2e = \Delta B \cdot S_{eff}$ is ~1.03 µm$^2$, corresponding to a circle with a diameter of 1.02 µm, which is in between the inner and outer diameter of the fabricated ring. This is similar with the results in device #1 and indicates that these periodic oscillations arise from the Little-Parks effect with charge 2*e*. We further measured the differential resistance at different magnetic fields and plot it as a colormap, as shown in Fig. 3c. Although the magnetoresistance curves of device #2 do not show multiple sets of oscillations like that observed in device #1, besides the Little-Parks oscillations with $\Delta B$ of 2.01 mT, the supercurrent interference pattern clearly shows another set of weak oscillations with a decaying height when increasing magnetic field. This set of oscillations does not influence the *R-B* curves, which might be because it is much weaker than the Little-Parks oscillations. But again, this observation demonstrates that the differential resistance colormap reveals more superconducting features than the *R-B* curves.

To further study the origin of the other set of weak oscillations, we investigated the intrinsic properties of the superconducting strip segment by using the 4-probe measurement (labeled as $I_{36}$-$V_{45}$) shown in the inset of Fig. 3d, which does not include the ring segment in the current source path. The *R-B* curves at different temperatures are shown in Fig. 3d. Notably, small oscillations are observed at low temperatures, which become weaker when temperature increases, as shown in the zoomed plot in Fig. 3e. Furthermore, the differential resistance colormap (Fig. 3f) reveals clear periodic oscillations with height gradually decaying when increasing magnetic field. There exhibit two sets of oscillations, which have similar central lobes and separated lobes with similar periods at higher magnetic fields. Each oscillation set has a slightly broader and higher central lobe than other lobes at high magnetic field. This is not the same as the standard Little-Parks oscillations where all lobes possess the same period [2,34]. It is not the same with the standard Fraunhofer pattern either, which has a high central lobe that is two times broader than other lobes [35,36]. These features of oscillations resemble the modified Fraunhofer pattern with non-uniform current distribution. The supercurrent interference pattern has also been observed in other pristine $KV_3Sb_5$ flakes, indicating it is an intrinsic property of $KV_3Sb_5$, a detailed analysis is included in the Supplementary Materials. In addition, the antisymmetric interference pattern in Fig. 3f indicates the presence of superconducting diode effect with different positive ($I_{c+}$) and negative ($I_{c-}$) critical current at a fixed magnetic field, (see detail discussion on the superconducting diode effect in Supplementary Materials).

We now compare the supercurrent interference pattern measured on the ring (Fig. 3c) and the strip itself (Fig. 3f). It's clear that the other set of weak oscillations of the ring is very similar to the intrinsic oscillations of the strip. This hints us the existence of the global critical current effect, which means that the critical current of a segment is not only determined by the measured segment, i.e. in between the voltage probes, but can also be influenced by other superconducting segments within the current source path, as reported in our previous work [37]. To further demonstrate the transfer of supercurrent interference between different segments, we measured the strip segment ($S_{45}$) while including the ring segment ($S_{23}$) in the current source path (source current between lead 1 and 6, defined as the global configuration) and plotted the differential resistance colormap shown in Fig. 4a. Interestingly, the colormap differs significantly from that in Fig. 3c (source current between lead 3 and 6 without including the ring on the path, defined as the local configuration). This indicates that in global configuration, the ring segment does influence the critical current behavior of the strip segment even when it is outside the voltage leads. For a better comparison of the critical current behaviors, we extracted the $I_{c+}$ from the colormaps shown in Fig. 3c, 3f and 4a and plotted them in Fig. 4b. Note that the weak oscillations in Fig. 3c were not extracted. We can clearly see that, for the strip segment, the critical current behaviors in global configuration ($I_{16}$-$V_{45}$) includes the oscillations from both ring and the strip. At lower magnetic field (between -1.5 mT and 1.5 mT), the supercurrent oscillations measured in global configuration are similar to that measured in local configuration, but with reduced height. As an example, the differential resistance as a function of bias current at 0 mT is normalized to their normal state resistance and plotted in Fig. 4c. However, at relatively high magnetic fields, the oscillation of the strip segment in global configuration ($I_{16}$-$V_{45}$) coincides very well with that of the ring segment ($I_{16}$-$V_{23}$), demonstrating the transfer of the Little-Parks oscillations from the ring segment to the strip segment. The normalized differential resistance as a function of bias current at 4 mT is shown in Fig. 4d, where the superconducting diode effect of the strip in local configuration ($I_{36}$-$V_{45}$) can be observed. However, when changing to the global configuration ($I_{16}$-$V_{45}$), both positive and negative critical currents align perfectly with those of the ring ($I_{16}$-$V_{23}$). Additionally, comparing the critical current of the strip in local configuration to global configuration, its $I_{c+}$ is decreased, whereas $I_{c-}$ is increased, showcasing the current-polarity-dependent global critical current effect.

**Discussion**

We now discuss the physical understanding of two phenomena observed in this work. One is the intrinsic supercurrent interference in a superconducting $KV_3Sb_5$ flake, and the other is the transfer of superconducting phase coherent phenomena.

The appearance of supercurrent interference pattern in a superconducting flake without fabricated ring or Josephson junction structures, indicates that the superconducting order parameters are not uniform in the material, and phase coherence happens in different superconducting regions. Intrinsic supercurrent interference is rare, but has been observed in some cases, including Weyl semimetal $MoTe_2$ with edge supercurrent from topological edge state [38], and superconductors with charge orders such as $CsV_3Sb_5$ [27,28], $Bi_2Sr_2CaCu_2O_{8+x}$ [39] and $TiSe_2$ [40], where the superconducting domains play the role for the phase coherence. The oscillations in $KV_3Sb_5$ flakes (Fig. 3f and results of device #3 in Supplementary Material) decay quickly after applying magnetic field, differing from the long-lived and slowly decayed supercurrent oscillations induced by topological edge states. Moreover, the effective area calculated from the oscillation period in several samples is also not consistent with the area either between voltage probes or current probes (See discussions in Supplementary Materials). These results indicate that the intrinsic supercurrent interference pattern in $KV_3Sb_5$ flakes is not induced by the topological edge state. Similar to $CsV_3Sb_5$, $KV_3Sb_5$ has been shown to possess time-reversal symmetry breaking charge orders coexisting with superconductivity, and chiral superconducting domains have been detected by scanning tunnelling microscopy [23]. In $CsV_3Sb_5$, both Little-Parks like oscillations and Fraunhofer patterns with thermal dynamic tunability were reported and believed to be related to the chiral superconductivity [27,28]. We also measured the supercurrent interference patterns under different thermal cycling, as the results shown in Supplementary Material. Different from the large thermal dynamic tunability observed in $CsV_3Sb_5$, only slight change of supercurrent oscillations was observed in $KV_3Sb_5$ and a cycling temperature above the critical temperature of charge density wave ($T_{CDW}$) seems to influence the oscillations more obviously. Although it's not clear if chiral superconductivity exists in $KV_3Sb_5$, the coherence between different superconducting domains should contribute to the observation of intrinsic superconducting phase coherence.

The transfer of supercurrent interference pattern in device #2 is a kind of global critical current effect, indicating the critical current of a superconducting segment is not only determined by its own microscopic structure but can also be influenced by adjacent segments within the current source path. Although at first glance, the global critical current effect seems to be explained by the extrinsic Joule heating effect, because that once a superconducting segment breaks superconductivity, it can generate heat due to the Joule heating effect, which may spreads to adjacent regions and drives them above their critical temperature. However, device #2 shows the current-polarity dependent global critical current effect (Fig. 4d), where $I_{c+}$ of $S_{45}$ is reduced but $I_{c-}$ of $S_{45}$ is increased, when changing from local configuration ($I_{36}$-$V_{45}$) to global configuration ($I_{16}$-$V_{45}$). The increase of $I_{c-}$ violates the Joule heating picture, which can only reduce the critical current by heating up the segment. The transfer of the Little-Parks oscillations (Fig. 4a and 4b) indicates that the phase winding generated in the ring propagates to adjacent region that exceeds the coherence length (tens of nm to the order of $10^2$ nm in $KV_3Sb_5$) of the Cooper pairs [15,41,42]. This also implies that the Cooper pairs can carry the phase coherence feature within the current source path. The study on such non-local phenomenon has just started, and more experiments such as the propagation length limit is essential to unveil the fundamental mechanism of the global critical current effect.

In summary, we observe rich oscillations in both magnetoresistance and supercurrent interference pattern in $KV_3Sb_5$ rings and flakes. A detailed comparison between different measurements reveals that magnetoresistance data alone is insufficient to fully capture the behavior of Cooper pairs in superconducting rings. Instead, characterization of the supercurrent interference pattern is essential for a comprehensive understanding of the superconducting phase coherence. Furthermore, we identify a global critical current effect associated with the transfer of superconducting interference between different segments of the sample. This demonstrates that the properties of the measured superconducting region can arise from other regions located outside the voltage leads, contradicting the conventional assumption of localized measurement. Although the microscopic origin of the global critical current effect remains unclear and requires further theoretical and experimental investigation, our results highlight the importance of accounting for the non-local transfer of superconducting interference when analyzing and interpreting superconducting transport measurements.

References

[1] A. A. Golubov, M. Yu. Kupriyanov, and E. Il'ichev, The current-phase relation in Josephson junctions, Rev. Mod. Phys. **76**, 411 (2004).
[2] W. A. Little and R. D. Parks, Observation of quantum periodicity in the transition temperature of a superconducting cylinder, Phys. Rev. Lett. **9**, 9 (1962).
[3] Y. Li, X. Xu, M.-H. Lee, M.-W. Chu, and C. L. Chien, Observation of half-quantum flux in the unconventional superconductor β-$Bi_2Pd$, Science **366**, 238 (2019).
[4] A. Almoalem, I. Feldman, I. Mangel, M. Shlafman, Y. E. Yaish, M. H. Fischer, M. Moshe, J. Ruhman, and A. Kanigel, The observation of π-shifts in the Little-Parks effect in 4Hb-$TaS_2$, Nat. Commun. **15**, 4623 (2024).
[5] E. Berg, E. Fradkin, and S. A. Kivelson, Charge-4e superconductivity from pair-density-wave order in certain high-temperature superconductors, Nat. Phys. **5**, 830 (2009).
[6] S. Zhou and Z. Wang, Chern fermi pocket, topological pair density wave, and charge-4e and charge-6e superconductivity in kagomé superconductors, Nat. Commun. **13**, 7288 (2022).
[7] S.-K. Jian, Y. Huang, and H. Yao, Charge- 4e superconductivity from nematic superconductors in two and three dimensions, Phys. Rev. Lett. **127**, 227001 (2021).
[8] D. F. Agterberg, Conventional and charge-six superfluids from melting hexagonal Fulde-Ferrell-Larkin-Ovchinnikov phases in two dimensions, Phys. Rev. B **84**, (2011).
[9] E. V. Herland, E. Babaev, and A. Sudbø, Phase transitions in a three dimensional U ( 1 ) × U ( 1 ) lattice London superconductor: Metallic superfluid and charge- 4 e superconducting states, Phys. Rev. B **82**, 134511 (2010).
[10] B. R. Ortiz et al., $CsV_3Sb_5$: A $Z_2$ Topological Kagome Metal with a Superconducting Ground State, Phys. Rev. Lett. **125**, 247002 (2020).
[11] B. R. Ortiz et al., New kagome prototype materials: discovery of $KV_3Sb_5$, $RbV_3Sb_5$, and $CsV_3Sb_5$, Phys. Rev. Mater. **3**, 094407 (2019).
[12] J.-X. Yin, B. Lian, and M. Z. Hasan, Topological kagome magnets and superconductors, Nature **612**, 647 (2022).
[13] K. Jiang, T. Wu, J.-X. Yin, Z. Wang, M. Z. Hasan, S. D. Wilson, X. Chen, and J. Hu, Kagome superconductors $AV_3Sb_5$ (A=K, Rb, Cs), Natl. Sci. Rev. **10**, nwac199 (2023).
[14] Y. Wang, H. Wu, G. T. McCandless, J. Y. Chan, and M. N. Ali, Quantum states and intertwining phases in kagome materials, Nat. Rev. Phys. **5**, 635 (2023).
[15] B. R. Ortiz, P. M. Sarte, E. M. Kenney, M. J. Graf, S. M. L. Teicher, R. Seshadri, and S. D. Wilson, Superconductivity in the Z2 kagome metal $KV_3Sb_5$, Phys. Rev. Mater. **5**, 034801 (2021).
[16] C. M. III et al., Time-reversal symmetry-breaking charge order in a kagome superconductor, Nature **602**, 245 (2022).
[17] H. Li, H. Zhao, B. R. Ortiz, T. Park, M. Ye, L. Balents, Z. Wang, S. D. Wilson, and I. Zeljkovic, Rotation symmetry breaking in the normal state of a kagome superconductor $KV_3Sb_5$, Nat. Phys. **18**, 265 (2022).
[18] H. Zhao, H. Li, B. R. Ortiz, S. M. L. Teicher, T. Park, M. Ye, Z. Wang, L. Balents, S. D. Wilson, and I. Zeljkovic, Cascade of correlated electron states in the kagome superconductor $CsV_3Sb_5$, Nature **599**, 216 (2021).
[19] X. Wu et al., Nature of Unconventional Pairing in the Kagome Superconductors $AV_3Sb_5$( A = K , Rb , Cs ), Phys. Rev. Lett. **127**, 177001 (2021).

[20] A. T. Rømer, S. Bhattacharyya, R. Valentí, M. H. Christensen, and B. M. Andersen, Superconductivity from repulsive interactions on the kagome lattice, Phys. Rev. B **106**, 174514 (2022).

[21] H. Chen et al., Roton pair density wave in a strong-coupling kagome superconductor, Nature **599**, 222 (2021).

[22] X.-Y. Yan et al., Chiral pair density waves with residual Fermi arcs in RbV$_3$Sb$_5$, Chin. Phys. Lett. **41**, 097401 (2024).

[23] H. Deng et al., Chiral kagome superconductivity modulations with residual Fermi arcs, Nature **632**, 775 (2024).

[24] D. F. Agterberg, J. C. S. Davis, S. D. Edkins, E. Fradkin, D. J. Van Harlingen, S. A. Kivelson, P. A. Lee, L. Radzihovsky, J. M. Tranquada, and Y. Wang, The physics of pair-density waves: cuprate superconductors and beyond, Annu. Rev. Condens. Matter Phys. **11**, 231 (2020).

[25] S.-L. Yu and J.-X. Li, Chiral superconducting phase and chiral spin-density-wave phase in a Hubbard model on the kagome lattice, Phys. Rev. B **85**, 144402 (2012).

[26] H. Deng et al., Evidence for time-reversal symmetry-breaking kagome superconductivity, Nat. Mater. **23**, 1639 (2024).

[27] T. Le et al., Superconducting diode effect and interference patterns in kagome CsV3Sb5, Nature **630**, 64 (2024).

[28] T. Le, Z. Xu, J. Liu, R. Zhan, Z. Wang, and X. Lin, Thermomodulated Intrinsic Josephson Effect in Kagome CsV$_3$Sb$_5$, Phys. Rev. Lett. **135**, 096002 (2025).

[29] J. Ge, P. Wang, Y. Xing, Q. Yin, A. Wang, J. Shen, H. Lei, Z. Wang, and J. Wang, Charge-4e and Charge-6e Flux Quantization and Higher Charge Superconductivity in Kagome Superconductor Ring Devices, Phys. Rev. X **14**, 021025 (2024).

[30] Chandra M. Varma and Ziqiang Wang, Extended superconducting fluctuation region and and flux quantization in a kagome compound with a normal state of order, Phys. Rev. B **109**, 219902 (2024).

[31] Y. Xie et al., Conventional superconductivity in the doped kagome superconductor Cs(V$_{0.86}$Ta$_{0.14}$)$_3$Sb$_5$ from vortex lattice studies, Nat. Commun. **15**, 6467 (2024).

[32] M. Roppongi et al., Bulk evidence of anisotropic s-wave pairing with no sign change in the kagome superconductor CsV$_3$Sb$_5$, Nat. Commun. **14**, 667 (2023).

[33] Y. Zhong et al., Nodeless electron pairing in CsV$_3$Sb$_5$-derived kagome superconductors, Nature **617**, 488 (2023).

[34] G. Kopnov, O. Cohen, M. Ovadia, K. H. Lee, C. C. Wong, and D. Shahar, Little-Parks Oscillations in an Insulator, Phys. Rev. Lett. **109**, 167002 (2012).

[35] M. Tinkham, *Introduction to Superconductivity* (Courier Corporation, 2004).

[36] B. D. Josephson, Possible new effects in superconductive tunnelling, Phys. Lett. **1**, 251 (1962).

[37] H. Wu, Y. Wang, and M. N. Ali, *The Global Critical Current Effect of Superconductivity*, arXiv:2412.20896.

[38] W. Wang, S. Kim, M. Liu, F. A. Cevallos, R. J. Cava, and N. P. Ong, Evidence for an edge supercurrent in the Weyl superconductor MoTe$_2$, Science **368**, 534 (2020).

[39] M. Liao, Y. Zhu, S. Hu, R. Zhong, J. Schneeloch, G. Gu, D. Zhang, and Q.-K. Xue, Little-Parks like oscillations in lightly doped cuprate superconductors, Nat. Commun. **13**, 1316 (2022).

[40] M. Liao, H. Wang, Y. Zhu, R. Shang, M. Rafique, L. Yang, H. Zhang, D. Zhang, and Q.-K. Xue, Coexistence of resistance oscillations and the anomalous metal phase in a lithium intercalated TiSe$_2$ superconductor, Nat. Commun. **12**, 5342 (2021).

[41] Y. Wang et al., Anisotropic proximity–induced superconductivity and edge supercurrent in Kagome metal, K$_{1-x}$V$_3$Sb$_5$, Sci. Adv. **9**, eadg7269 (2023).

[42] S. Ni et al., Anisotropic Superconducting Properties of Kagome Metal CsV$_3$Sb$_5$, Chin. Phys. Lett. **38**, 057403 (2021).

**Data availability**

The data that support the findings of this study are available from the corresponding author upon reasonable request.

**Acknowledgments:**

H.W. acknowledges that this research was supported by from NWO Talent Programme VENI financed by the Dutch Research Council (NWO) VI. Veni.222.380. Y.W. acknowledges the support from NWO Talent Programme VENI financed by the NWO, project no. VI.Veni.212.146, and KU Leuven Special Research Fund STG/24/071 and no.3E250622. M.N.A acknowledges support from the NWO Talent Programme VIDI financed by the NWO VI.Vidi.223.089, the Kavli Institute Innovation Award 2023, the Kavli Institute of Nanoscience Delft, and the research program "Materials for the Quantum Age" (QuMat, registration number 024.005.006) which is part of the Gravitation program financed by the Dutch Ministry of Education, Culture and Science (OCW). B.R.O. gratefully acknowledges support from the U.S. Department of Energy (DOE), Office of Science, Basic Energy Sciences (BES), Materials Sciences and Engineering Division. S.D.W. gratefully acknowledges support via the UC Santa Barbara NSF Quantum Foundry funded via the Q-AMASE-i program under award DMR-1906325.

**Author contributions:** Y.W. and H.W. conceived and designed the study. B.R.O and S.D.W. grew the single crystal. H.W., Y.W. M.D., H.M.H and fabricated the devices. H.W. and Y.W. performed all the measurements. H.W. and Y.W. carried out the data analysis. M.N.A is the Principal Investigator. All authors contributed to the preparation of manuscript.

**Competing interests:** The authors declare that they have no competing interests.

**Additional information**

**Correspondence and requests for materials** should be addressed to H.W. and Y.W.

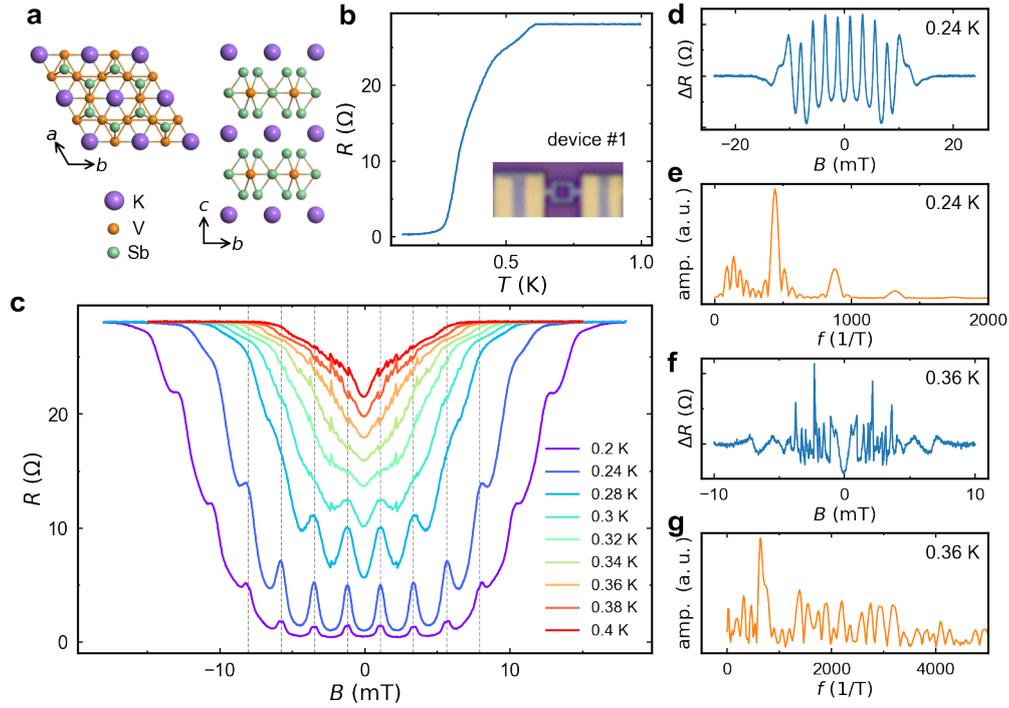

**Fig 1. Little-Parks oscillations of a KV3Sb5 ring device (device #1). a**. Crystal structure of the kagome material $KV_3Sb_5$. **b**. *R-T* curve of the $KV_3Sb_5$ ring. **c**. *R-B* curves of the $KV_3Sb_5$ ring at different temperatures. The period of the Little-Parks oscillations at low temperatures is indicated by the grey dotted lines with a spacing of 2.3 mT. However, the periodicity becomes complicated and messy at higher temperatures. **d**. *R-B* curve at 2.4 K after removing its background. **e**. FFT result based on **d**, where the dominated peak at 439 (1/T) can be clearly observed, indicating the periodicity of the Little-Parks oscillations. The peaks observed at higher frequencies are the higher harmonics of the dominant peak. **f**. *R-B* curve at 3.6 K after removing its background. **g**. FFT result based on **f**, where lots of peaks with comparable amplitudes can be observed, indicating the oscillations do not originate from the Little-Parks effect. The backgrounds of **d** and **f** are obtained using Savitzky-Golay filter on the original curve with a window length of 1051 and 431, respectively, and a polynomial order of 3.

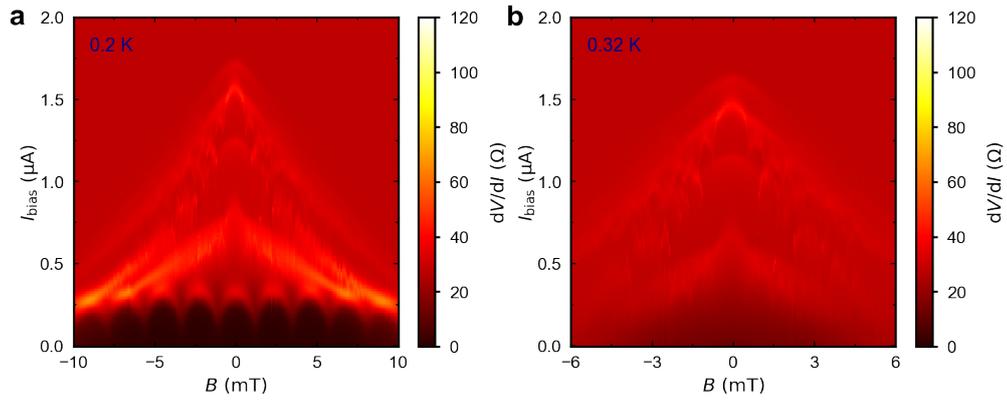

**Fig 2. Differential resistance colormaps of KV$_3$Sb$_5$ ring (device #1). a**. Differential resistance colormap measured at 0.2K, where a set of Little Parks oscillations can be clearly observed at low bias currents. The anomalous oscillations on top of a diamond shaped background can be observed at higher bias currents. **b**. Differential resistance colormap measured at 0.32 K, where the set of Little-Parks oscillations can no longer be observed, but the anomalous oscillations remain.

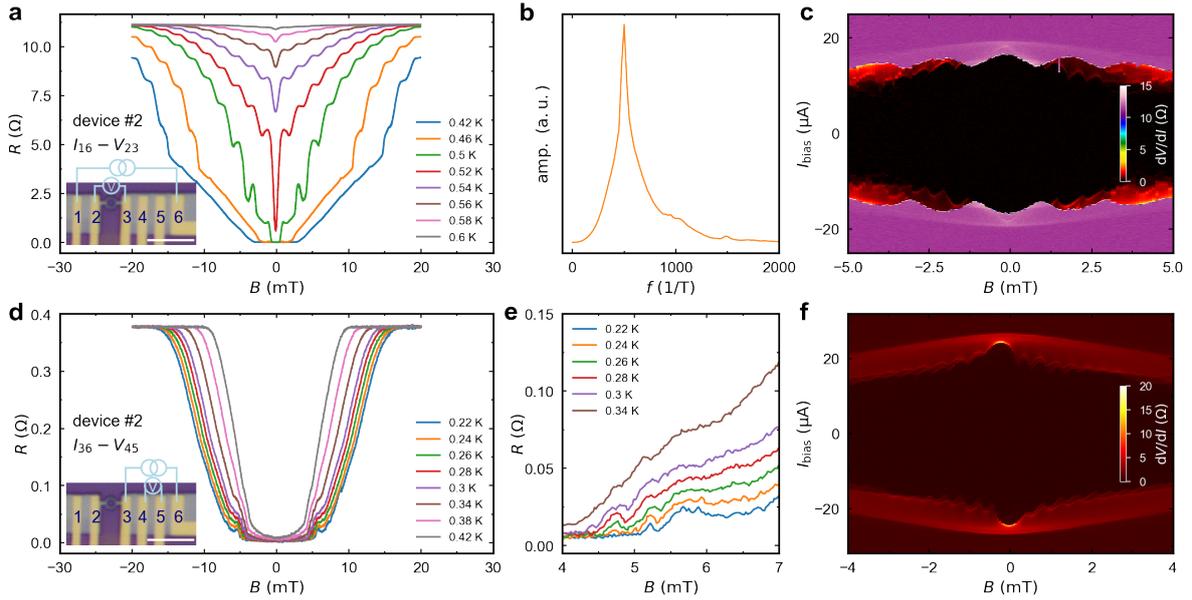

**Fig 3. Oscillations in another KV3Sb5 ring device (device #2)**. **a**. *R-B* curves at different temperatures of the ring, where the Little-Parks oscillations can be clearly observed. The inset is the optical image of the KV$_3$Sb$_5$ ring device and the measurement schematics. The scale bar is 5 μm. **b**. FFT of the *R-B* curve measured at 0.52 K, where the dominant peak at 501 (1/T) can be observed. **c**. Differential resistance colormap as a function of magnetic field and bias current. The Little Parks oscillations can be clearly observed, but another set of oscillations can also be seen. **d**. *R-B* curves at different temperatures of the strip segment ($S_{45}$), where another set of oscillations can be seen. **e**. The zoomed-in plot of the oscillations in **d**, the oscillations become weaker when increasing temperature. **f**. colormap of the differential resistance as a function of magnetic field and bias current. The Fraunhofer-like oscillation can be clearly observed in the flake region of the device.

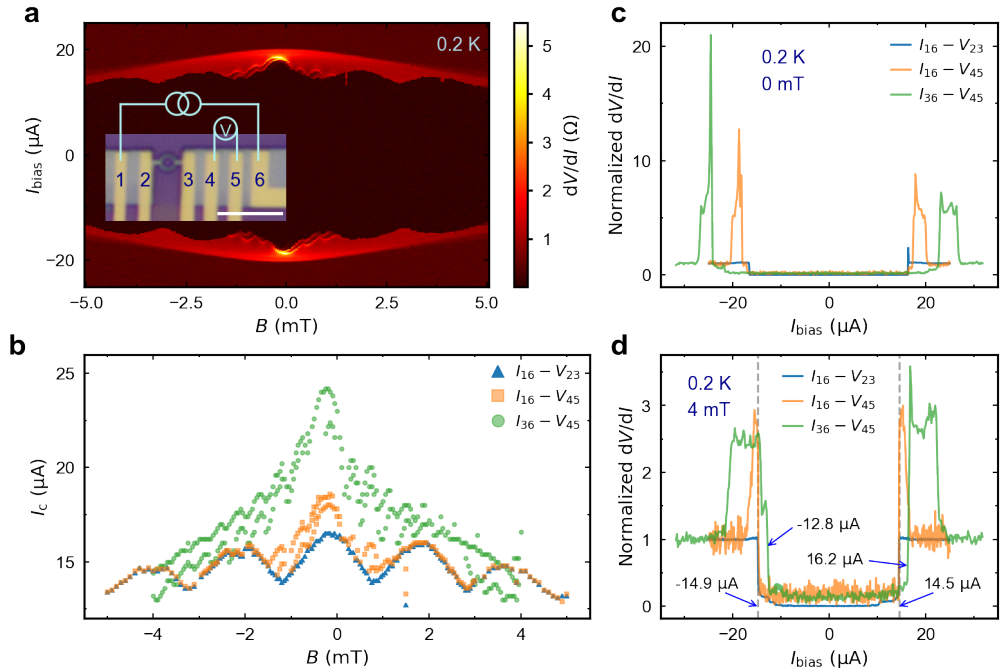

**Fig 4. Global critical current effect in the KV$_3$Sb$_5$ ring-flake device (device #2). a**. Colormap of the differential resistance as a function of magnetic field and bias current of the strip region while including the ring segment in the current source path. The inset shows the measurement diagram. **b**. The comparison of extracted critical currents of the ring and flake segments in different configurations. **c**. The comparison of the normalized differential resistance of the ring and the flake region at 0.2 K and 0 mT in different configurations. The differential resistances are normalized with their normal state resistance. The differential resistance is measured from 0 to positive bias current and 0 to negative bias current, so both sides are critical currents, and no return currents are shown here. **d**. The comparison of the normalized differential resistance of the ring and the flake region at 0.2 K and 4 mT with different measurement configurations.